\documentclass[reprint,amsmath,amssymb,aps,floatfix]{revtex4-1}

\usepackage{graphicx}
\usepackage[ansinew]{inputenc}
\usepackage{dcolumn}
\usepackage[]{units}	
\usepackage{xcolor}
\usepackage{soul}
\usepackage{changes}
\usepackage{hyperref}
\usepackage[all]{hypcap}
\usepackage{textcomp}
\usepackage{upgreek}
\usepackage{enumerate}

\begin{document}

\title{Optoelectrical cooling of polar molecules to sub-millikelvin temperatures}

\author{Alexander Prehn}
\author{Martin Ibr\"ugger}
\author{Rosa Gl\"ockner}
\author{Gerhard Rempe}
\author{Martin Zeppenfeld}
	\email{Martin.Zeppenfeld@mpq.mpg.de}
\affiliation{Max-Planck-Institut f\"ur Quantenoptik, Hans-Kopfermann-Strasse 1, 85748 Garching, Germany}%

\date{\today}

\begin{abstract}
We demonstrate direct cooling of gaseous formaldehyde (H$_2$CO) to the microkelvin regime. Our approach, optoelectrical Sisyphus cooling, provides a simple dissipative cooling method applicable to electrically trapped dipolar molecules. By reducing the temperature by three orders of magnitude and increasing the phase-space density by a factor of $\sim10^4$ we generate an ensemble of $3\cdot 10^5$ molecules with a temperature of about 420\,\textmu K, populating a single rotational state with more than 80\,\% purity. 
\end{abstract}

\maketitle

The ability to cool atoms to ultracold temperatures has led to previously unimagined applications ranging from metrology~\cite{Huang2013} to the simulation of quantum many-body systems~\cite{Bloch2008}. Cooling molecules to ultracold temperatures promises a similar variety of applications, including precision spectroscopy~\cite{Truppe2013,Schnell2011}, investigation of fundamental physics~\cite{Baron2014,DeMille2008}, ultracold chemistry~\cite{Bell2009,Stuhl2014}, study of highly anisotropic quantum gases~\cite{Baranov2012,Wall2015}, and quantum information~\cite{Andre2006,Wei2011}. Research with ultracold atoms has been enabled by a simple and robust technique, i.e., laser cooling. Despite substantial progress in slowing and cooling molecules directly~\cite{Quintero-Perez2014,Barry2014,Patterson2015,Narevicius2012,VandeMeerakker2012,Chervenkov2014,Lu2014,Merz2012,Marx2013,Hogan2009} and in synthesizing diatomic molecules from ultracold alkali atoms~\cite{Ni2008}, a similarly versatile method to cool molecules to ultracold ($T<1\mathrm{mK}$) temperatures has been lacking. 

An ideal cooling scheme for molecules should satisfy the following criteria: First, the technique should be simple so that it can be applied to different experiments. Second, it should be robust, without the need of permanent maintenance. Third, it should be applicable to a large variety of molecule species of interest. Fourth, and most important, it needs to achieve temperatures and molecule numbers which are useful for further experiments. 

Here we demonstrate that optoelectrical Sisyphus cooling~\cite{Zeppenfeld2009} satisfies all these criteria: First, it requires only a single laser, a single microwave and a single radio frequency (RF) source. It also requires a novel electrostatic trap~\cite{Englert2011}, but this consists mainly of two microstructured electrode plates which can trivially be reproduced with modern lithographic techniques. Second, it is robust as it is routinely operated 24 hours a day in our laboratory. Third, cooling as first demonstrated with methyl fluoride (CH$_3$F)~\cite{Zeppenfeld2012a} is now applied to formaldehyde (H$_2$CO) for which interesting ultracold collision experiments have been proposed~\cite{Hudson2006}. Fourth, about $3\cdot10^5$ molecules can now be cooled to 0.4mK. Although no fundamental cooling limit has been reached, two important goals are achieved: a record-large ensemble of ultracold molecules has been prepared, and a temperature has been reached which is so low that further experiments become possible, like the operation of a molecular fountain~\cite{Bethlem2008} or the use of microwave~\cite{DeMille2004} and optical dipole traps.

Optoelectrical Sisyphus cooling proceeds in an electric trap which produces a box-like potential for each low-field-seeking molecular state. Specifically, the trap features a homogeneous offset field in a large part of the volume and strongly increasing fields at the edges. Kinetic energy is removed by allowing molecules to repeatedly move up the electric field gradient of the trapping potential in rotational states with strong Stark interaction, and back down in states with weaker Stark interaction. We implement cooling in a closed scheme of trapped rovibrational states exploiting the spontaneous decay of the $v_1$ C-H stretch vibrational mode for dissipation. States are labeled with vibrational quantum number $v$ and symmetric-top rotational quantum numbers $J$, $K$, and $M$ as $|v{;}J{,}{\mp}K{,}{\pm}M\rangle$ with ${\mp}K$ chosen positive~\cite{Suppl}. We use the rotational states characterized by $J=3,4$, $|K|=3$. Although formaldehyde is a slightly asymmetric rotor molecule, moderately strong electric fields couple inversion-doublet rotational states such that our states essentially possess the properties of symmetric-top states~\cite{Suppl}.

\begin{figure}[tb]
\centering
\includegraphics{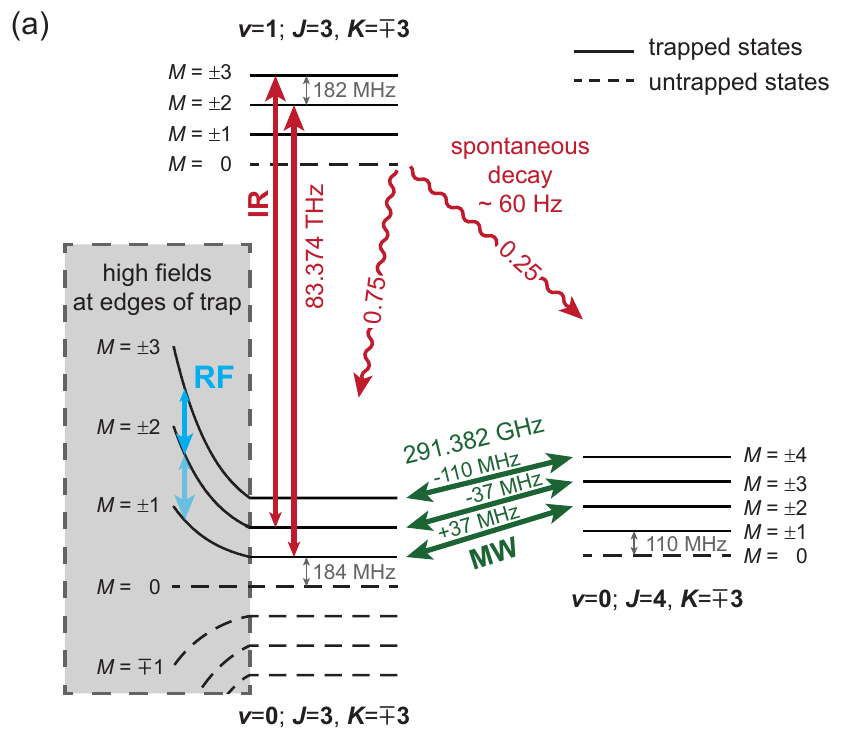}
\includegraphics{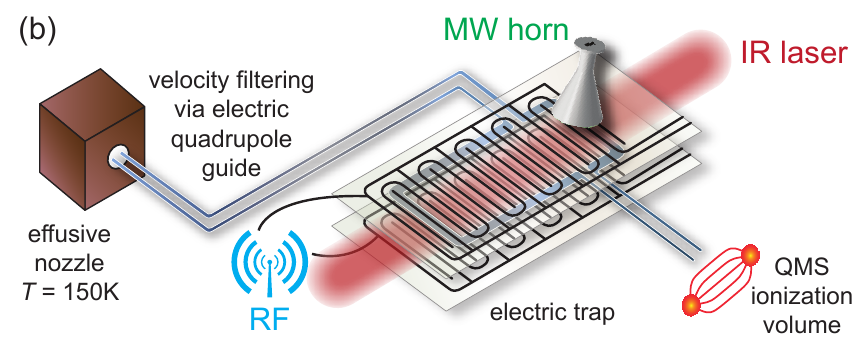}
\caption{
(a)~Level scheme for optoelectrical Sisyphus cooling as explained in the main text. Radiation (solid arrows) couples rovibrational states, wavy arrows indicate spontaneous decay channels and their branching ratio (without resolving the $M$-substates).
The Stark splitting of neighboring $M$ sublevels and transition frequencies~\cite{Cornet1980,Rothman2009} are given for the homogeneous-field region of the trap ($V_{\mathrm{trap}}{=}{\pm}1500\mathrm{V}$).
(b)~Experimental setup. Molecules are loaded into the trap from a velocity-filtered thermal source ($T\approx150\mathrm{K}$) via an electric quadrupole guide~\cite{Junglen2004a,Englert2011}. The radiation fields needed for cooling are applied as indicated. A second guide brings molecules to a quadrupole mass spectrometer (QMS), where a time-resolved count rate is recorded.
}
\label{fig:Fig1}
\end{figure}

The level scheme for cooling is shown in Fig.~\ref{fig:Fig1}(a). Exciting a $\Delta M{=}{+}1$ infrared (IR) transition from states $|0{;}3{,}3{,}M{<}3\rangle$ and coupling the states $|0{;}3{,}3{,}M{>}0\rangle$ and $|0{;}4{,}3{,}M{+}1\rangle$ with microwave (MW) radiation results in optical pumping of molecules to the strongly trapped highest $M$ sublevels. This is based on spontaneous decay from the excited states obeying the selection rules $\Delta J,\Delta M=0,\pm1$ and $\Delta K=0$~\cite{Suppl}. Coupling neighboring $M$ sublevels with RF in the strong-field edge region of the trap closes the cooling cycle: transitions to more weakly trapped states in high electric field remove kinetic energy. Losses to untrapped states are minimized by driving these transitions at a rate slow compared to the optical pumping rate to strongly trapped states~\cite{Zeppenfeld2012a}. The spontaneous decay rate of $\sim$60Hz limits the optical pumping and hence the speed of cooling. Here we profit from formaldehyde having a four times faster vibrational decay than the previously used species methyl fluoride. Note that the RF acts on all molecules, although the potential energy curves and transitions in the edge region of the trap are only sketched for one set of states in the figure. 

The experimental setup is shown and explained in Fig.~\ref{fig:Fig1}(b). Its key part is the electrostatic trap where molecules are trapped between a pair of microstructured capacitor plates (spaced 3mm) and a surrounding electrode~\cite{Englert2011}. Alternating high voltages ${\pm}V_{\mathrm{trap}}$, applied to adjacent electrodes of the microstructure, produce strong trapping fields. To suppress so-called Majorana flips to untrapped states and to spectroscopically separate the rotational $M$-substates, an offset electric field is created by applying additional voltages ${\pm}V_{\mathrm{offset}}$ across the capacitor. This yields a homogeneous offset field in the center of the trap, with a finite roughness owing to ${\pm}V_{\mathrm{trap}}$~\cite{Englert2011}. Except for detection of molecules, ${\pm}V_{\mathrm{offset}}$ always equals $5\%$ of ${\pm}V_{\mathrm{trap}}$, thus ensuring a fixed relation of offset and trapping fields. The unique design offers long storage times for polar molecules. Uncooled molecules can be stored with a 1/e decay time on the order of 10s~\cite{Englert2011}, while for cooled molecules it can be as long as a minute~\cite{Suppl}.

\begin{figure*}[tb]
\centering
\includegraphics{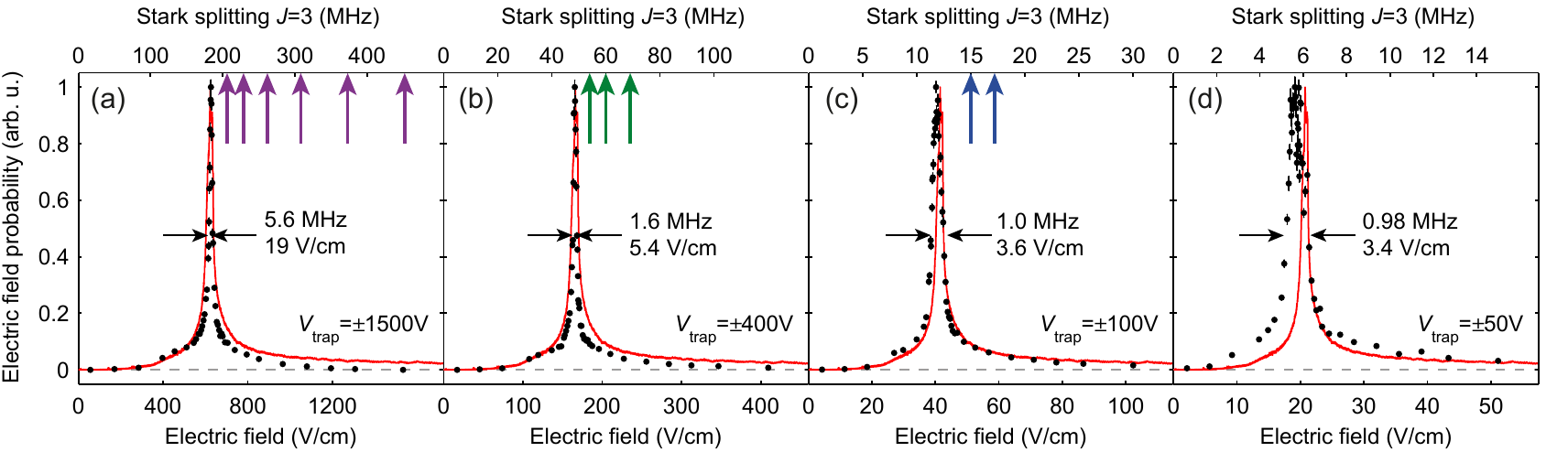}
\caption{
Measured (dots) and simulated (solid line, see supplement~\cite{Suppl}) electric-field distribution in the trap, both normalized to the maximum. Error bars represent the 1$\sigma$ statistical error. We indicate the FWHM of the measured distributions. The applied trap voltages were varied over a factor of 30 (see lower right corner of panels). The upper horizontal axes show the Stark splitting of the states $|0{;}3{,}3{,}M\rangle$ in frequency units. Configurations (a)-(c) are used during cooling and arrows indicate the frequencies of the RF coupling ($f_{\mathrm{RF}}$) applied sequentially for cooling. Panels (c), (d) are relevant for the determination of the final kinetic energy.
}
\label{fig:fielddist}
\end{figure*}

The actual shape of the trap potential strongly influences both the cooling sequence and the determination of the final temperature. For example, we later subtract the offset potential energy from the measured total energy of the molecules to determine their kinetic energy based on the assumption of a box-like potential. This simple picture is only valid if the kinetic energy of the molecules is sufficiently larger than the roughness of the offset field of the electric box. Therefore, we investigate the potential landscape in detail by measuring electric-field distributions. Evidently, a box-like potential with a large homogeneous-field region translates to a strongly peaked electric-field distribution with the width of the peak giving the roughness of the offset. The distribution is measured by performing Stark spectroscopy on the single MW transition coupling the states $|0{;}3{,}3{,}3\rangle$ and $|0{;}4{,}3{,}4\rangle$, similarly to previous work~\cite{Glockner2015}. The line shape of the measured depletion spectrum is primarily given by Stark broadening and thus allows us to extract the probability for a specific electric field to occur in the trap~\cite{Glockner2015}.

The measured and simulated electric-field distributions for the four trap voltage configurations used for the experiments in this paper are plotted in Fig.~\ref{fig:fielddist}. Configurations (a)-(c) are used during cooling, whereas panels (c) and (d) are relevant for the determination of the final temperature. Narrow peaks confirm that the electric fields are indeed homogeneous in a large fraction of the trap volume. The simulation~\cite{Suppl} predicts the peak position and the overall shape of the distribution quite well. In a perfect trap, the field distributions are expected to scale with the applied trap voltage. However, the relative widths of the distributions increase and a slight shift of the peak is observable for small voltages. In particular, halving the voltage from $V_{\mathrm{trap}}{=}{\pm}100\mathrm{V}$ to $V_{\mathrm{trap}}{=}{\pm}50\mathrm{V}$ (Fig.~\ref{fig:fielddist}(c), (d)) still approximately halves the strength of the homogeneous field, but does not narrow the width noticeably any more. We attribute this effect to the existence of surface charges on the microstructured capacitor plates~\cite{Suppl}. Note that the measured distribution is convolved with the probability for molecules to be at a given potential energy, which mainly influences the high-field tail of the distribution~\cite{Glockner2015}.

The aspects of the box-like potential which are relevant to this work can be captured by two parameters of the measured electric-field distributions. First, the center of the peak defines the homogeneous offset electric field. This allows calculation of the Stark splitting of $M$ sublevels in this region as $f_{\mathrm{offset}}=\frac{\mathcal{E}\mu}{h}\cdot\frac{K}{J(J+1)}$, with electric field strength $\mathcal{E}$, electric dipole moment $\mu$ and the rotational quantum numbers (cf. upper horizontal axes of Fig.~\ref{fig:fielddist}). Consequently, $M\cdot f_{\mathrm{offset}}$ defines the potential energy offset, which we later subtract, for each low-field seeking state. Second, the width of the field distribution compared to the kinetic energy of the molecules defines whether we can treat the potential as box-like in the first place. Molecules loaded into the trap initially can reach electric fields above 10kV/cm and the potential is then clearly box-like.

The previous results can now be used to discuss the effect of the RF on cooling. The amount of kinetic energy extracted per induced RF transition is given by the applied RF frequency $f_{\mathrm{RF}}$, and by $f_{\mathrm{offset}}$ as $\Delta E \sim h\left(f_{\mathrm{RF}}-f_{\mathrm{offset}}\right)$. On one hand, a larger $f_{\mathrm{RF}}$ leads to more cooling per transition. On the other hand, the molecules have to possess sufficient kinetic energy to reach the high electric fields where the RF is resonant to allow for a transition. Consequently, $f_{\mathrm{RF}}$ has to be reduced as cooling advances. We chose to lower $\left(f_{\mathrm{RF}}-f_{\mathrm{offset}}\right)$ stepwise in factors of $\sim\sqrt{2}$ every 2s corresponding to about one induced transition per applied frequency~\cite{Zeppenfeld2012a}.

In the course of cooling $f_{\mathrm{RF}}$ approaches $f_{\mathrm{offset}}$ and $\left(f_{\mathrm{RF}}-f_{\mathrm{offset}}\right)$ becomes comparable to the width of the electric-field distribution. This can be seen in Fig.~\ref{fig:fielddist}(a), where vertical arrows denote the last six $f_{\mathrm{RF}}$  applied with the initial trap-voltage configuration. Thus, the molecules do not move in a well-defined box potential any more and spend significant time in not very well-defined isolated regions of low electric field. To maintain a simple box-like potential we ramp down the trap voltages adiabatically twice during cooling to $V_{\mathrm{trap}}{=}{\pm}400\mathrm{V}$ and $V_{\mathrm{trap}}{=}{\pm}100\mathrm{V}$. This shifts the offset of the potential, $f_{\mathrm{offset}}$, and reduces the width of the field distribution as intended. We perform in total five more cooling steps in a reduced trap potential (arrows in Fig.~\ref{fig:fielddist}(b), (c)).

\begin{figure}[tb]
\centering
\includegraphics{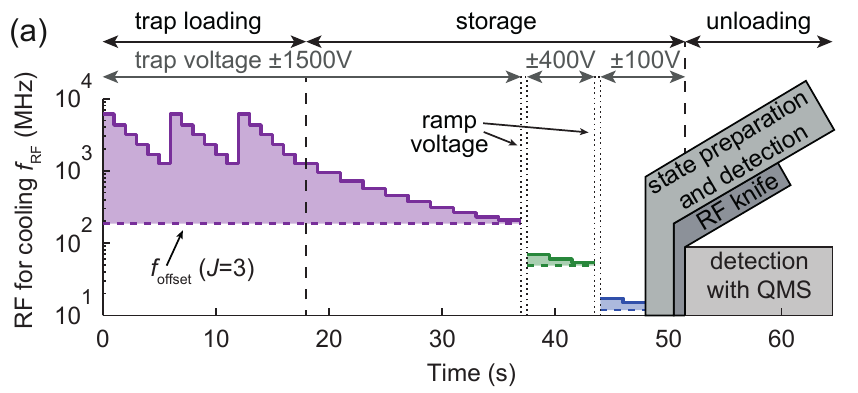}
\includegraphics{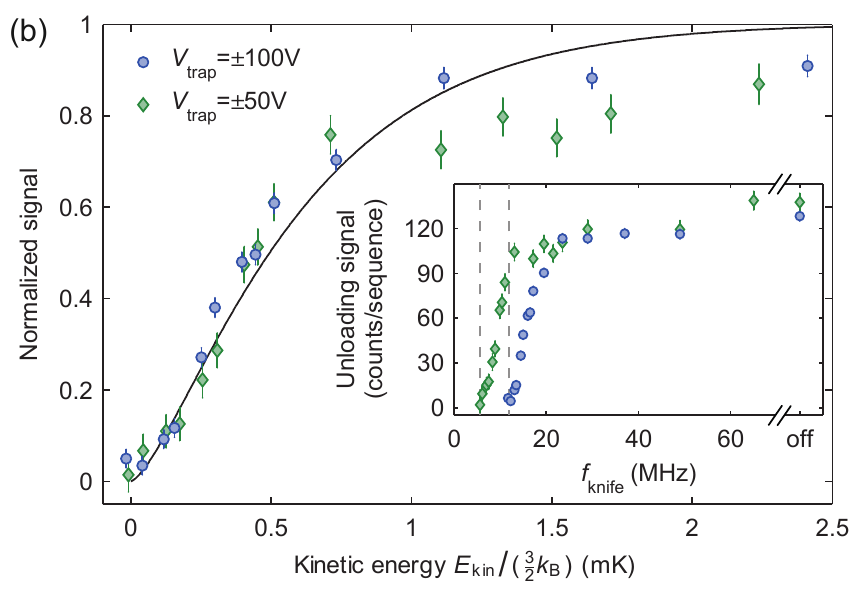}
\caption{
(a)~Experimental sequence. The RF applied for cooling is plotted together with the trapping sequence vs. time. The dashed horizontal lines mark $f_{\mathrm{offset}}$ for $J{=}3$. The experimental steps performed after cooling are explained in the main text.
(b)~Measurement of the kinetic energy via RF knife-edge filters. The inset shows the signal of molecules remaining in the states $|0{;}3{,}3{,}3\rangle$ versus $f_{\mathrm{knife}}$ with the filter applied with two distinct trap voltages. The dashed lines mark the potential energy offset $f_{\mathrm{offset}}$ which is 11.9MHz (5.7MHz) for $V_{\mathrm{trap}}{=}{\pm}100\mathrm{V}$ ($V_{\mathrm{trap}}{=}{\pm}50\mathrm{V}$). 
The main panel plots the same data versus kinetic energy, normalized to the signal without RF knife and on a narrower horizontal scale.
The solid line represents an integrated Boltzmann distribution for a temperature of 420\textmu K. Error bars denote the 1$\sigma$ statistical error.
}
\label{fig:cooling}
\end{figure}

The preceding discussion leads to the following sequence to cool and detect a sample of molecules, consisting of six parts as depicted in Fig.~\ref{fig:cooling}(a). First, molecules are continuously loaded into the trap (voltage $\pm1500\mathrm{V}$) for 18s with cooling already applied. 
The lifetime of molecules in the trap increases substantially for colder molecules~\cite{Zeppenfeld2012a}. Therefore, cooling as many hot molecules entering the trap at different times and with varying energy as fast as possible to an energy with a decent lifetime is desirable. To achieve this, we apply and cycle six $f_{\mathrm{RF}}$ during loading. Then, the ensemble is stored and further cooled while reducing $f_{\mathrm{RF}}$ and lowering the trapping potential as described above. After cooling, the molecules are prepared in a single rotational state. Thus, the trap potential is identical for all observed molecules. Specifically, we optically pump to the state $|0{;}3{,}3{,}3\rangle$~\cite{Glockner2015a,Suppl} and utilize a rotational-state-detection technique~\cite{Glockner2015,Suppl}. Next, the energy of the molecular ensemble is probed by applying a strong RF field that acts as a knife-edge filter and only eliminates hotter molecules (see below). Finally, the molecules are unloaded from the trap and counted with a quadrupole mass spectrometer (QMS). Presently, very cold molecules can only be extracted efficiently from the trap if the ensemble is parametrically heated before unloading~\cite{Suppl}. The reason for heating is not fundamental but technical: possibly due to disturbed electric fields created by surface charges or a misalignment at the transition from the trap to the guide slow molecules are lost before they reach the detector~\cite{Suppl}.

The RF knife-edge filter employed for measuring the energy of the molecules works as follows. A strong RF field drives $\Delta M=-1,\Delta K=0$ transitions to the untrapped $M=0$ states and quickly depletes all those molecules from the trap possessing sufficient energy to reach the (higher) electric field where the RF is resonant. Hence, an RF knife with a frequency $f_{\mathrm{knife}}$ truncates the energy distribution of the molecules at a known total energy and leaves only less energetic molecules in the trap. Scanning $f_{\mathrm{knife}}$ and observing the molecule signal thus yields an integrated energy distribution of the trapped molecules. Note that due to formaldehyde being an asymmetric rotor additionally $\pm K\rightarrow \mp K$ transitions can be induced in lower electric field~\cite{Suppl}. These transitions require orders of magnitude higher RF power and are thus well-separated from the desired energy-dependent filtering~\cite{Suppl}.

The inset of Fig.~\ref{fig:cooling}(b) shows the remaining signal in the state $|0{;}3{,}3{,}3\rangle$ versus $f_{\mathrm{knife}}$. To address the issue of offset subtraction, we measure the energy distribution for two distinct trap voltages: $V_{\mathrm{trap}}{=}{\pm}100\mathrm{V}$, the trap potential present in the final two cooling steps, and $V_{\mathrm{trap}}{=}{\pm}50\mathrm{V}$. 
In both cases, a knife with $f_{\mathrm{knife}}\approx f_{\mathrm{offset}}$, resonant to the offset of the box-like potential, depletes all molecules from the trap as expected. With rising knife frequency we observe steeply rising flanks which are clearly separated due to the different potential energy offsets. At higher frequencies a slight further increase towards the data point without RF knife is visible. Those molecules, about 10\% of the ensemble, were not cooled efficiently in the last cooling steps and therefore have a much higher kinetic energy than the vast majority of the fully cooled ones. If necessary, this high-energy part of the molecular ensemble could be removed from the trap by applying a suitable RF knife.

To obtain a kinetic energy distribution from the measurement, the contribution of potential energy has to be considered. As we can treat the trap potential as box-like, we account for this by subtracting the offset potential energy, $M \cdot h \cdot f_{\mathrm{offset}}$, extracted from the measured electric-field distributions (see above, Fig.~\ref{fig:fielddist}(c), (d)). We verified that systematic errors caused by this simple approach are smaller than our statistical uncertainty. Consequently, we find for the kinetic energy $E_{\mathrm{kin}} = M \cdot h\left(f_{\mathrm{knife}} - f_{\mathrm{offset}}\right)$. The measured data with the offset potential energy subtracted is shown in the main panel of Fig.~\ref{fig:cooling}(b). Normalized to the signal without RF knife, the two curves obtained with different trap voltages show a good overlap. This is expected from the fact that the subtraction of a well-defined offset potential energy should not influence the kinetic energy distribution. 

From the data we compute a median kinetic energy of $E_{\mathrm{kin}} / \left(\frac{3}{2}k_{\mathrm{B}}\right) = \left(420 \pm 90\right)$\textmu K with the factor of 3 accounting for the three translational degrees of freedom in a box potential. The energy at which half of the molecular ensemble is depleted was determined by fitting the curve for $V_{\mathrm{trap}}{=}{\pm}100\mathrm{V}$ with a linear slope in the vicinity of this kinetic energy. As a comparison we additionally plot a Boltzmann distribution for a temperature of 420\textmu K. The good agreement of thermal distribution and measured data supports our interpretation of $E_{\mathrm{kin}}{/}\left(\frac{3}{2}k_{\mathrm{B}}\right)$ as an approximate temperature.

The produced molecular ensemble is both large and internally pure. Calibrating the sensitivity of the QMS, we determine the number of cooled molecules unloaded from the trap to be $3 \cdot 10^5$, accurate to within a factor of two~\cite{Suppl}. We measure $(83 \pm 3)\%$ of the molecules to populate the single rotational state $|0{;}3{,}3{,}3\rangle$~\cite{Suppl}. The result can be compared to an uncooled reference ensemble: molecules unloaded from the trap after 18s of trap loading and two seconds of storage without any manipulation, resulting in $10^6$ molecules with 460mK in the states $|0{;}3{,}3{,}M\rangle$. This comparison yields a reduction of kinetic energy by a factor of 1000 and an increase in phase-space density by about $10^4$.

With the simple and robust technique of optoelectrical Sisyphus cooling we produced a large ensemble of trapped ultracold molecules. In principle, further cooling with the same method is possible, if the technical issue with the surface charges, which broaden the electric fields in the present setup, is solved. Additionally, larger ensembles of molecules could be produced by loading buffer-gas cooled~\cite{VanBuuren2009} and decelerated~\cite{Chervenkov2014} molecules into our trap. We note that our method relies on rather general properties of polar molecules and should thus be applicable to a wide range of additional species~\cite{Zeppenfeld2009}. 

The temperature and ensemble size reached enable further experiments. The low velocity of the cooled molecules of $ \sim 0.6\mathrm{m/s}$ makes fountain experiments feasible~\cite{Bethlem2008}. The ability to control the final kinetic energy and rotational state is an ideal starting point for collision studies and investigation of cold and ultracold chemistry~\cite{Bell2009,Stuhl2014}. Finally, the temperature achieved should allow efficient transfer to a microwave~\cite{DeMille2004} or optical trap where molecules can be held in their absolute ground state, a prerequisite for sympathetic~\cite{Tokunaga2011,Lutz2014} or evaporative cooling.

Similar results on direct cooling of molecules to sub-millikelvin temperatures using a radio-frequency magneto-optical trap are reported in Ref.~\cite{Norrgard2016}.

\clearpage

\renewcommand{\thefigure}{S\arabic{figure}}
\renewcommand{\theHfigure}{S\arabic{figure}} 
\setcounter{figure}{0}

\section*{Supplemental Material}

In this supplement we summarize the following additional information: We discuss the rotational structure of formaldehyde in Sec.~\ref{formaldehyde} and how it affects the optimization of the RF knife-edge filters in Sec.~\ref{rfknife}. Sec.~\ref{details} sums up experimental details like the production of formaldehyde and the generation of the radiation fields. An exemplary measurement of the lifetime of cooled molecules in the electric trap is shown in Sec.~\ref{lifetime}. Further, we discuss rotational-state detection (Sec.~\ref{rsd}) and optical pumping to the state $|0;3,3,3\rangle$ (Sec.~\ref{op}). The simulation of the electric-field distribution is described in Sec.~\ref{simu}. Finally, we detail our observations suggesting the existence of surface charges on the microstructured electrode array of the trap (Sec.~\ref{charges}), discuss the unloading of molecules from the trap (Sec.~\ref{unloading}), and summarize the deduction of the number of cooled molecules (Sec.~\ref{density}).

\subsection{The slightly asymmetric rotor formaldehyde}
\label{formaldehyde}

Formaldehyde is a slightly asymmetric prolate top molecule~\cite{Townes1975}. In this section we will argue that we can treat the sets of rotational states chosen for the experiment as symmetric-top rotational states for most aspects of our experiment.

Using asymmetric rotor notation, the general form of the wave function of a prolate asymmetric top is~\cite{Townes1975}
\begin{equation*}
|J,K_A,K_C\rangle = \sum_{K=K_A\pm2n}{a_{J,K} \Psi_{J,K}}
\end{equation*}
with $\Psi_{J,K}$ being symmetric-top wave functions and $a_{J,K}$ suitable expansion coefficients. An asymmetric rotor state $|J,K_A,K_C\rangle$ is characterized by the total angluar momentum $J$ and the $K$ values of the limiting prolate ($K_A$) or oblate ($K_C$) symmetric top cases.

The three sets of rotational states addressed throughout the experiments (the $J=5$ state playing a role in state preparation and detection) are $|J{=}3,K_A{=}3,K_C{=}\left\{0,1\right\}\rangle$, $|J{=}4,K_A{=}3,K_C{=}\left\{1,2\right\}\rangle$, and $|J{=}5,K_A{=}3,K_C{=}\left\{2,3\right\}\rangle$. Due to the slightly asymmetric structure there is a small inversion splitting between states with equal $J,K_A$ but different $K_C$. These splittings are \unit[0.66]{MHz}, \unit[4.6]{MHz}, and \unit[18]{MHz}, respectively~\cite{Cornet1980}, for our three sets of states. In an electric field, as present in our trap, the inversion-split states couple strongly such that it essentially holds $|J,K_A,K_C\rangle\approx\Psi_{J,K=K_A}$ and contributions from other $\Psi_{J,K'}$ in the expansion above are negligible~\cite{Townes1975}. Consequently, these states behave like symmetric-top states. In particular, this guarantees that trapped, low-field-seeking vibrational excited states predominantly decay back to low-field-seeking ground states. Further, there is essentially a linear Stark shift ensuring strong electric trapping. With a numerical diagonalization of the molecule Hamiltonian including electric-field interaction we find that this holds for electric fields greater than about \unit[1]{V/cm}, \unit[10]{V/cm}, and \unit[40]{V/cm} for the three sets of states. Therefore, we can treat formaldehyde as a symmetric top molecule in most of our experiments. 

For state nomenclature (as explained in the main paper) we thus only use $K=K_A$, the quantum number of a prolate symmetric top, and omit $K_C$. Using this notation, we describe the trapped, low-field seeking states with $|v;J,\mp K,\pm M\rangle$ with $\mp K$ chosen positive.

\begin{figure}[tb]
\centering
\includegraphics{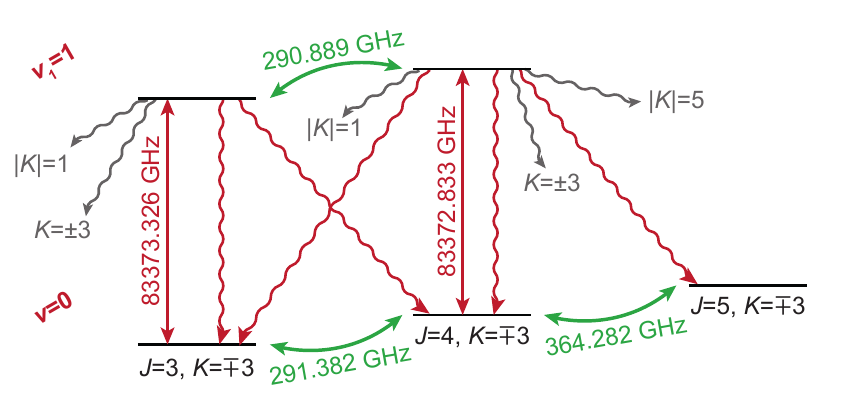}
\caption{
Level scheme showing all transitions addressed during the experiments (without $M$-substructure). Long wavy arrows indicate the dominant spontaneous decay channels, whereas the short, gray wavy arrows indicate theoretically possible but strongly suppressed loss channels due to the slightly asymmetric rotor structure. The manifold $|0;5,3,M\rangle$ plays a role for rotational-state preparation and detection.
}
\label{fig:schemeH2CO}
\end{figure}

The selection rules for electric dipole transitions are $\Delta J,\Delta M=0,\pm1$. Additionally, the selection rule for symmetric top molecules $\Delta K=0$ holds for formaldehyde in most cases. Transitions with $\Delta K=\pm2,\pm4,...$ (occuring due to the slightly asymmetric structure) are suppressed by at least a factor of $10^2$~\cite{Cornet1980}. We have to take into account such transitions only when optimizing the strong RF radiation used for the knife-edge filter (see Sec.~\ref{rfknife}).

Figure~\ref{fig:schemeH2CO} shows a level scheme of the used rotational states and frequencies of all addressed transitions. Precise rotational transition frequencies are available in the literature~\cite{Cornet1980}. For cooling and optical pumping we excite the $v_1$ vibrational C-H stretch mode at a wavelength of about \unit[3.6]{\textmu m}. Vibrational line assignments and frequency values published in the HITRAN database~\cite{Rothman2009} are verified with saturation spectroscopy with sub-MHz accuracy using a multi-pass Herriot cell. The excited states $|1;3,3,M\rangle$ decay with a rate of $\sim$\,\unit[60]{Hz} to $|0;3,3,M\rangle$ and $|0;4,3,M\rangle$~\cite{Rothman2009}.

\subsection{RF knife-edge filters}
\label{rfknife}

\begin{figure}[tbp]
\centering
\includegraphics{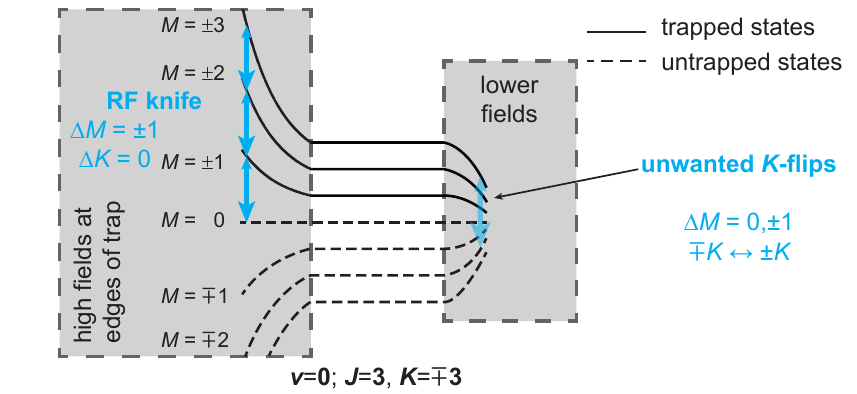}
\caption{
Level scheme for RF knife-edge filters.
The Stark shifted energy levels of the states $|0;3,3,M\rangle$ are shown for different electric-field regions in the trap (unshaded area represents the homogeneous electric field). RF applied with a specific frequency induces allowed transitions in the edge region of the trap, quickly depleting hot molecules which are able to reach the higher electric fields. This process acts as the RF knife. Additionally, $K$-flip transitions can occur in regions with lower electric field leading to unwanted depletion of hot \textit{and} cold molecules. The latter process is forbidden in first approximation and thus requires dramatically larger RF power.
}
\label{fig:scheme_RFknife}
\end{figure}

The use of RF radiation as a tool to measure the kinetic energy of a molecular ensemble in the trap relies on the fact that RF can couple successive $M$ sublevels of rotational states and induce transitions to untrapped states. Molecules reaching the $M{\leq}0$ states are lost from the trap almost immediately. The frequency $f_{\mathrm{knife}}$ can be chosen such that molecules need a certain energy to reach the regions of high electric field where the RF is resonant. This is illustrated in the left part of Fig.~\ref{fig:scheme_RFknife}. As all molecules populate the same internal state $|0;3,3,3\rangle$ initially, a sufficiently strong RF pulse truncates the energy distribution of the ensemble at a defined value by depleting all hotter molecules from the trap.

Due to formaldehyde being a slightly asymmetric rotor the wave function for the state with $J=-K=K_A=3$ contains contributions with $K=+3$ and $K=\mp1$ (see Sec.~\ref{formaldehyde}). Therefore, a non-zero matrix element exists for transitions from $\mp K$ to $\pm K$ (with $\Delta M=0,\pm1$), which are strictly forbidden for a rigid symmetric rotor. A strong RF coupling with frequency $f_{\mathrm{knife}}$ can induce depletion via such transitions. These $K$-flip transitions occur in lower electric fields where $f_{\mathrm{knife}}$ equals five or six times the splitting of $M$ sublevels (Fig.~\ref{fig:scheme_RFknife}). Therefore, it commonly depletes molecules independent of their energy. However, the matrix element is small compared to the one for $\Delta K=0$ for our experimental parameters allowing separation of the two processes. Note that the states with $J=3$ and $|K|=1$ of formaldehyde are more than a THz away and transitions to those states can therefore not be induced with the frequencies we apply.

\begin{figure}[tbp]
\centering
\includegraphics{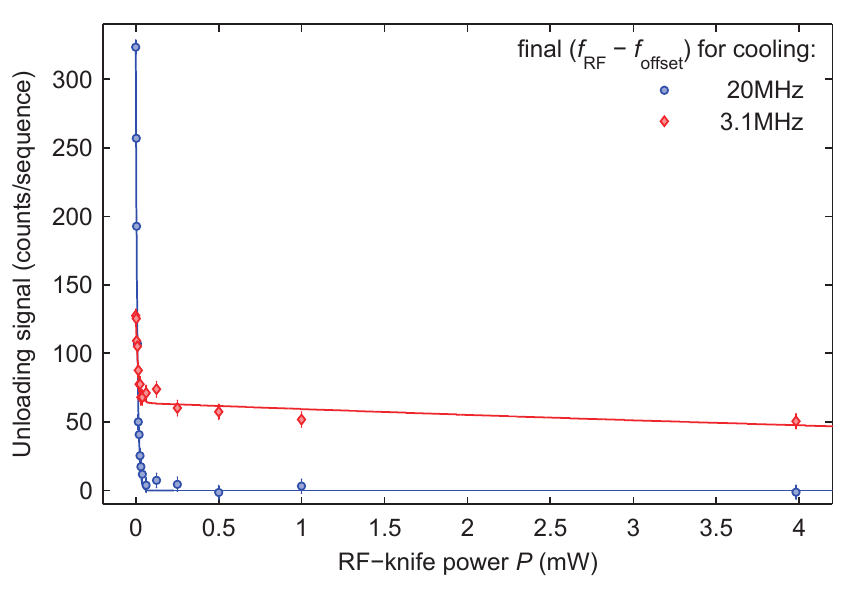}
\caption{
Scan of the RF power applied as a knife-edge filter ($f_{\mathrm{knife}}=16.5\,\mathrm{MHz}$) for two cooling sequences.
The experimental sequence is the one explained in the main text, except that cooling was stopped earlier for one of the two scans. The data for the hotter ensemble is well-fitted with an exponential model as shown because all molecules are depleted by the RF knife. The scan for the colder ensemble is fitted with a double-exponential model to account for the additional decay of signal at high RF powers due to induced $K$-flip transitions. Error bars show the $1\sigma$ statistical error.
}
\label{fig:RFknife_power}
\end{figure}

The largely different matrix elements for the wanted transitions of the RF knife and the unwanted $K$-flip transitions manifest themselves, when the RF power used for the knife-edge filter is scanned. We perform such a power scan for two experimental sequences producing molecular ensembles with about a factor of four difference in kinetic energy (full sequence and cooling stopped at $f_{\mathrm{RF}}-f_{\mathrm{offset}}=20\,\mathrm{MHz}$, see Fig.~\ref{fig:RFknife_power}). A knife-edge filter with $f_{\mathrm{knife}}=16.5\,\mathrm{MHz}$ is applied at a trap voltage of $\pm100\,\mathrm{V}$ ($f_{\mathrm{offset}}=11.9\,\mathrm{MHz}$). The hotter ensemble is depleted completely with low power filters. In the colder ensemble, depletion of molecules occurs with two power scales. The rapid decay at low powers is due to the RF knife. Depletion at high RF power, on the other hand, is caused by $K$-flip transitions. Such measurements allow us to choose an adequate power for an RF knife-edge filter which saturates the low-power process, but does not lead to significant depletion via $K$-flips. An adequately chosen filter hence depletes molecules depending on their energy.

For the measurement presented in the main paper we chose the following procedure. For every knife frequency $f_{\mathrm{knife}}$ we performed an RF power scan with a molecular ensemble having a suitable kinetic energy to allow both depletion processes to be resolved. The appropriate power for a knife-edge filter is then chosen such that the low-power process is clearly saturated (power at least five times the fitted decay constant) but depletion due to $K$-flips is still suppressed. Note that during cooling $K$-flip transitions are not an issue, because the RF is applied with dramatically reduced power.

\subsection{Details of the experiment}
\label{details}

This section summarizes the production of gaseous formaldehyde, the maximum electric fields applied for trapping and the generation of the radiation fields needed for cooling.

Formaldehyde in the gas phase is produced as described in Ref.~\cite{Motsch2009} by heating para-formaldehyde powder to a temperature around \unit[85]{$^\circ$C}. Unwanted water and polymer rests are removed from the gas by conducting it through a dry-ice cold trap at a temperature of $-78\,^\circ\mathrm{C}$. Then, the formaldehyde is injected into the vacuum and the quadrupole guide via a ceramic nozzle which is kept at a temperature of about \unit[150]{K}.

The large rotational constants of formaldehyde and sufficient vapor pressure down to almost \unit[140]{K} allow efficient velocity filtering with the electric quadrupole guide~\cite{Junglen2004a}. During loading the confining electric field applied to the quadrupole guide is \unit[30]{kV/cm}, whereas the trap voltage of $\pm1500\,\mathrm{V}$ used in the first part of the cooling sequence results in a nominal trap depth of \unit[50]{kV/cm}~\cite{Zeppenfeld2013}. Approximately 20\,\% of all molecules entering the trap in low-field-seeking states populate the states used for cooling, $|0;3,3,M\rangle$ and $|0;4,3,M\rangle$.

IR radiation to excite the $v_1$ vibrational band (C-H stretch mode) of formaldehyde is produced by a CW optical parametric oscillator referenced to a frequency comb. The IR beam illuminates a large fraction of the trap, driving the addressed vibrational transitions with a rate on the order of \unit[1]{kHz}. For optical pumping to the state $|0;3,3,3\rangle$ it is required to drive several vibrational transitions at the same time (see Sec.~\ref{op}). A quasi-simultaneous driving is achieved by changing the frequency of the IR source in a few milliseconds and cycling through the desired transition frequencies.

At present, we can generate MW around $290-310\,\mathrm{GHz}$ to couple the rotational states $J=3$ and $J=4$ as well as around $355-375\,\mathrm{GHz}$ to couple $J=4$ and $J=5$ in formaldehyde (cf. Fig.~\ref{fig:schemeH2CO}). Up to about \unit[20]{mW} of MW power can be generated in each frequency band with two independent amplifier-multiplier chains which are seeded by a single MW frequency synthesizer. The ability to switch between different frequencies in a few microseconds allows us to efficiently couple many rotational $M$ sublevels at essentially the same time, which is required for, e.g., rotational-state detection.

RF to couple neighboring $M$ sublevels for cooling and the knife-edge filter is applied directly to the contact leads of the trap microstructure. Lower frequencies are coupled in easily, for frequencies higher than \unit[1]{GHz} naturally occurring electric resonances are exploited. Due to these resonances occurring in the entire spectrum the applied RF power has to be optimized for every used frequency~\cite{Zeppenfeld2012a}.

\subsection{Lifetime of molecules in the trap}
\label{lifetime}

\begin{figure}[tbp]
\centering
\includegraphics{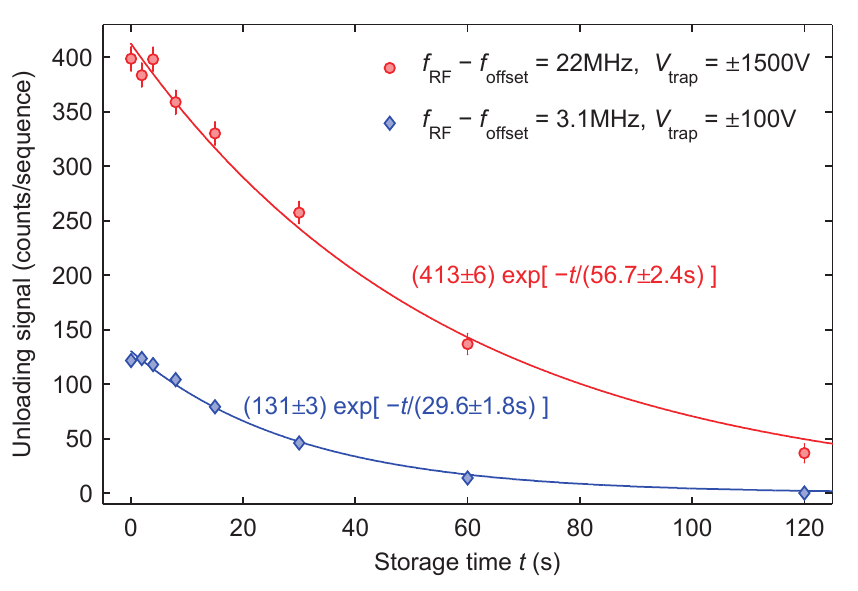}
\caption{
Lifetime of molecules in the trap. 
Molecules are cooled (cooling stopped at the specified value of $f_{\mathrm{RF}}-f_{\mathrm{offset}}$), prepared in the state $|0;3,3,3\rangle$, and stored in the trap for a varying amount of time with the same trap voltage as was applied during the final cooling step. This voltage is $15$ times smaller for the colder ensemble. We plot the state-selected signal for $|0;3,3,M\rangle$. Both curves are fitted with an exponential decay. Error bars denote the $1\sigma$ statistical error.
}
\label{fig:lifetime}
\end{figure}

For the successful implementation of optoelectrical cooling, a long lifetime of molecules in the trap is essential. Here, the trap lifetime depends on both the applied trap voltages, specifically the trap depth as well as the offset electric field, and on the energy of the molecules~\cite{Zeppenfeld2013}. For the maximum trap voltage of $\pm1500\,\mathrm{V}$, the hottest uncooled molecules are lost from the trap with a $1/e$ decay time of less than \unit[1]{s} whereas for colder uncooled molecules this increases to about \unit[10]{s}~\cite{Englert2011}. These numbers can be further improved by cooling the molecules~\cite{Zeppenfeld2012a}.

Figure~\ref{fig:lifetime} shows the remaining molecule signal versus storage time for two exemplary cooling sequences. First, we investigate the lifetime of the fully cooled ensemble (experimental sequence detailed in the main text, final $f_{\mathrm{RF}}-f_{\mathrm{offset}}=3.1\,\mathrm{MHz}$), which is stored with a voltage of $V_{\mathrm{trap}}{=}{\pm}100\,\mathrm{V}$, the voltage configuration of the final cooling steps. This yields a decay constant of \unit[30]{s}. This number improves if the trap voltage is increased. To demonstrate this, we prepare a slightly hotter ensemble (with a shorter cooling sequence, final $f_{\mathrm{RF}}-f_{\mathrm{offset}}=22\,\mathrm{MHz}$) and store the molecules with the maximum trap voltage of $\pm1500\,\mathrm{V}$. Then, we observe a 1/e decay time of almost a minute.

\subsection{Rotational-state detection}
\label{rsd}

Our detector, the QMS, is not sensitive to the internal state of the molecules. For state detection we selectively remove sets of rotational states from the trap by coupling them to untrapped states before unloading the remaining background of molecules. The difference between the total signal without removal and the background constitutes the state-selective unloading signal. We use MW radiation to deplete molecules populating certain states from the trap and therefore call the method microwave depletion (MWD). This is described in great detail in Ref.~\cite{Glockner2015}. In this section we summarize only its application to formaldehyde and the implementation in the present experiment.

Using MWD, state-sensitive depletion is performed by coupling trapped $M$ sublevels of at least two neighboring rotational states $J$ to the respective untrapped $M=0$ states. The population in the addressed states, which are coupled with $\Delta M=\pm1$ transitions, is mixed and molecules reaching the $M=0$ states get lost almost immediately. The level scheme in Fig.~\ref{fig:schemeH2CO} shows the sets of states which we can couple with the present setup.

In this work, we show state-selected signals for two manifolds of rotational states: $|0;3,3,M\rangle$ and $|0;J,3,M\rangle$ for $J=3,4$. For the former we subtract the signal measured with MWD applied for $J=3,4,5$ ($|K|=3$) from the one with MWD for $J=4,5$ ($|K|=3$). The latter is measured by subtracting the signal of MWD for $J=3,4$ ($|K|=3$) from the signal of all trapped molecules.
For MWD, a microwave pulse is applied consisting of various frequencies which are chosen such that rotational transitions are driven resonantly in the homogeneous-field region of the trap. 
MWD eliminates all molecules residing in the addressed states in about a second.

After cooling we prepare the molecular ensemble in the state $|0;3,3,3\rangle$. To measure the rotational-state purity after preparation, an $M$ sublevel selective state detection procedure is needed. Therefore, we determine the fraction of molecules populating $|0;3,3,3\rangle$ with respect to the total signal in $|0;3,3,M\rangle$ in the following way. MWD for $J=4,5$ ($|K|=3$) is applied with two additional microwaves coupling $|0;3,3,2\rangle$ and $|0;3,3,1\rangle$ to $|0;4,3,M\rangle$. This is compared to MWD for $J=3,4,5$ ($|K|=3$). The difference of these two signals yields the signal resulting from the state $|0;3,3,3\rangle$.

\subsection{Optical pumping to $|0;3,3,3\rangle$}
\label{op}

The potentials in our electrostatic trap depend on the rotational state. Notably for the measurement of the kinetic energy distribution via RF knife-edge filters we require the molecules to be in a well-defined potential and hence a single rotational state. Therefore, we optically pump the ensemble to the state $|0;3,3,3\rangle$ after cooling. Similar to Sisyphus cooling and previous work~\cite{Glockner2015a}, optical pumping via the vibrational mode is used in conjunction with MW coupling of rotational states. In particular, we drive three IR transitions quasi-simultaneously: $|0;3,3,M\rangle \leftrightarrow |1;3,3,M{+}1\rangle$, $|0;4,3,M\rangle \leftrightarrow |1;4,3,M{+}1\rangle$, and (with reduced duty cycle) $|0;4,3,M\rangle \leftrightarrow |1;4,3,M\rangle$. To achieve this, the IR frequency is quickly ramped between the three mentioned transitions every \unit[10]{ms}, \unit[10]{ms}, and \unit[5]{ms}, respectively. Furthermore, we couple $|0;4,3,M{>}0\rangle \leftrightarrow |0;5,3,M{+}1\rangle$ in the ground state and $|1;3,3,M{>}0\rangle \leftrightarrow |1;4,3,M{+}1\rangle$ in the vibrational excited state with MW (see Fig.~\ref{fig:schemeH2CO} for transition frequencies). As a result, the state $|0;3,3,3\rangle$ is the only dark state in the manifold of addressed vibrational ground states $|0;J,3,M\rangle$ for $J=3,4,5$. \unit[1.5]{s} of optical pumping at a trap voltage of $\pm100\,\mathrm{V}$ are sufficient to accumulate most of the molecules in the target state.

With this scheme cold molecules can be pumped to $|0;3,3,3\rangle$ with practically no losses and high efficiency. By measuring the rotational-state distribution after optical pumping (experimental sequence of main paper) we find that $\left(86\pm2\right)\,\%$ of the total signal originates from the states $|0;3,3,M\rangle$. Some molecules remain in $|0;4,3,M\rangle$ and $|0;5,3,M\rangle$, whereas $\left(5\pm1\right)\,\%$ are in states which are not addressed by the scheme at all. $\left(96\pm3\right)\,\%$ of the molecules in $|0;3,3,M\rangle$ populate the state $|0;3,3,3\rangle$. In total, this yields the specified rotational-state purity of $\left(83\pm3\right)\,\%$ for the molecular ensemble prepared in $|0;3,3,3\rangle$. The losses caused by optical pumping, about $\left(2\pm1\right)\,\%$, are negligible.

\subsection{Simulation of the electric-field distribution}
\label{simu}

In the main text, we compare the measured electric-field distribution to a simulation. To this end, the electric fields were simulated on a grid (10\,\textmu m $\times$ 10\,\textmu m $\times$ 20\,\textmu m grid spacing) considering the entire three-dimensional trap geometry. The simulation predicts the peak position, i.e., the value of the homogeneous offset field, and the shape of the distribution quite well indicating a good understanding of the trap electric fields. However, for trap voltages $|V_{\mathrm{trap}}|\geq400\,\mathrm{V}$ the measured peak is even narrower than the simulation. We attribute this deviation to numerical inaccuracies of the simulation.

\subsection{Surface charges on the electrode microstructure of the trap}
\label{charges}

For small trap voltages ($|V_{\mathrm{trap}}|\leq100\,\mathrm{V}$) the electric-field distributions presented in the main paper are substantially broader than expected and the peak is slightly shifted. We attribute both effects to the existence of surface charges on the microstructured capacitor plates. In this section we discuss additional observations backing up this interpretation.

The capacitor plates constituting the trap consist of an array of microstructured chromium electrodes on a glass substrate fabricated via optical lithography~\cite{Englert2011}. The plates currently installed in the experiment are completely coated with the polymer Cyclotene. This allows much higher voltages to be applied to the electrodes without flashovers than with uncoated versions of the microstructure and, thus, increases the maximally reached trap depths. However, due to the reduced electrical conductivity of the polymer, this additional layer might be the reason for the accumulation of surface charges.

Besides the altered electric-field distributions these surface charges manifest themselves in a second observation: if we switch off the trapping voltages completely for a certain time, molecules survive in the trapping volume for a time approximately an order of magnitude longer than what is expected considering their velocity. Although it is not fully understood how these charges accumulate on the surface of the trap, we could eliminate them in earlier experiments by heating up the entire vacuum chamber with the trap to $200\,^\circ\mathrm{C}$. After the bake-out the decay of molecule signal while switching off the trap was again as fast as expected. This suggests that it will be possible to understand the origin and finally eliminate the creation of surface charges in the future.

\subsection{Efficient unloading of molecules from the trap}
\label{unloading}

\begin{figure}[t]
\centering
\includegraphics{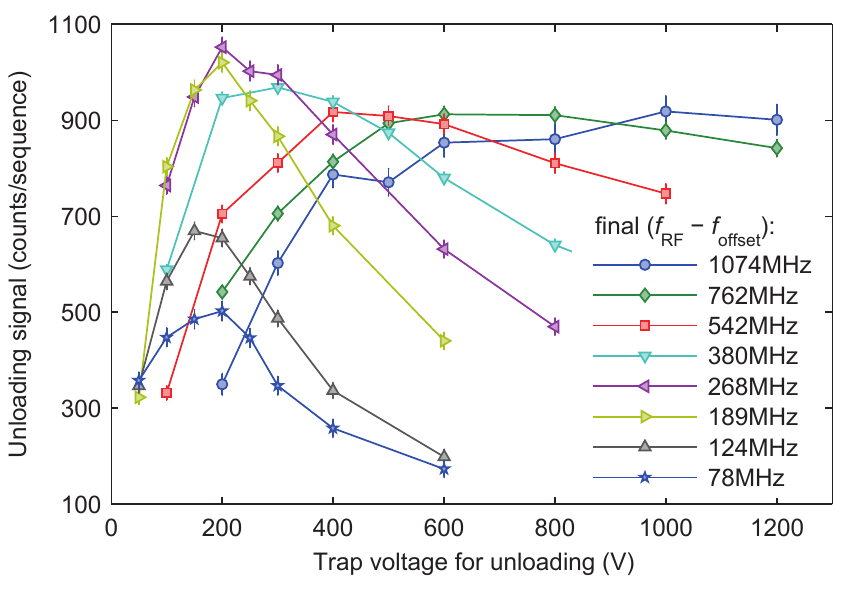}
\caption{
Unloading molecules from the trap.
Molecule signal versus trap voltage applied during unloading for various cooling sequences. Here, the molecules were not prepared in $|0;3,3,3\rangle$. Instead, we measured the signal of molecules populating the states characterized by $J=3,4$ ($|K|=3$). 
Solid lines are guides to the eye, keys specify when cooling was stopped, and all error bars represent the 1$\sigma$ statistical error.
}
\label{fig:unloading}
\end{figure}

In order to efficiently extract the molecules from the trap and guide them to the ionization volume of the QMS, the electric fields in the trap and the quadrupole guide have to be adjusted to the kinetic energy of the molecules (and to the Stark shift of the rotational state or set of states)~\cite{Zeppenfeld2012a,Glockner2015}. On the one hand, a small confining electric field strength leads to losses of hotter molecules. On the other hand, a higher electric field strength reduces the probability for colder molecules to find the trap exit hole to the QMS. 

To optimize the trap voltage applied during unloading we usually measure the integrated unloading signal with varying voltages applied. For unloading we apply a higher offset electric field between the capacitor plates, i.e., $\pm V_{\mathrm{offset}}=20\,\% |V_{\mathrm{trap}}|$, because we found that during molecule extraction (as opposed to storage) a higher offset electric field increases the molecule signal.

Data for optimizing the unloading voltage for a few cooling sequences is shown in Fig.~\ref{fig:unloading}. 
For hotter molecules (cooling stopped at higher frequency $f_{\mathrm{RF}}-f_{\mathrm{offset}}$) the optimal unloading voltage shifts towards lower values as the molecules are cooled further, satisfying the expectation. In particular, both the left and right flanks of the curves are shifted. For the three coldest ensembles, however, where the optimal unloading voltage based on scaling should be considerably smaller than \unit[200]{V}, only the right flank of the curve shifts and we observe a dramatic decrease in signal. Apparently, the unloading efficiency drops abruptly if molecules are cooled further. Possible reasons were already discussed in the main text.

Efficient unloading and detection of colder ensembles in the trap is possible by parametrically heating the molecules directly before trap unloading, using the following sequence. Rapidly ramping to an increased offset electric field in one half of the trap~\cite{Englert2011} with the other half remaining at the usual offset allows molecules residing in the region with higher field to roll down an electric field gradient and be accelerated. After a hold time of \unit[50]{ms}, the configuration of offset electric fields in the two halves of the trap is reversed. In total, we switch the offset electric fields 20 times during \unit[1]{s}, with the ramps performed in about \unit[1]{ms}. Molecules which were sufficiently cold initially thereby gain enough kinetic energy to be optimally unloaded from the trap at a voltage of $\pm200\,\mathrm{V}$. With parametric heating applied, a dramatic loss in signal for a final cooling frequency of $\left(f_{\mathrm{RF}}-f_{\mathrm{offset}}\right)<189\,\mathrm{MHz}$ is no longer observed, showing that the losses without heating are caused during unloading and not during cooling. 

A final interesting feature in Fig.~\ref{fig:unloading} is the fact that from $\left(f_{\mathrm{RF}}-f_{\mathrm{offset}}\right)=1074\,\mathrm{MHz}$ to $\left(f_{\mathrm{RF}}-f_{\mathrm{offset}}\right)=268\,\mathrm{MHz}$, we observe an increase in signal for optimal unloading voltage. This is observed despite an expected decrease in molecule number due to losses during additional cooling. However, the increased signal can be explained by a higher detection efficiency in the QMS for slower molecules since the ionization probability is proportional to the inverse velocity.

\subsection{Number of cooled molecules}
\label{density}

We calibrate the number of molecules exiting the trap to the quadrupole mass spectrometer (QMS) for detection as follows. Due to the low ionization probability for a given molecule traversing the ionization region of the QMS, the QMS count rate $s$ is proportional to the density of molecules in the ionization region $\rho$, independent of molecule velocity, i.e. $s=C\cdot\rho$. The coefficient $C$ quantifies the sensitivity of the QMS. The number of molecules $N$ exiting the trap is then equal to
\begin{equation}
\label{QMS calibration}
N=S\cdot A\cdot v\cdot \frac{1}{C}.
\end{equation}
Here, $S$ is the integrated background subtracted signal (counts) measured by the QMS during unloading, $A$ is the area of the molecular beam at the QMS, and $v$ is the velocity of molecules in the QMS.

The quantities in Eq.~\ref{QMS calibration} are determined as follows. The integrated signal $S$ is measured directly and the beam area $A$ is measured by varying the position of the QMS. The velocity $v$ is extracted from the median energy of molecules in the trap measured via RF knife-edge filters, based on the fact that the potential energy of molecules is entirely converted to kinetic energy before molecules enter the QMS. All these quantities can be measured relatively accurately.

The greatest difficulty in measuring the number of molecules exiting the trap originates from the need to determine the sensitivity coefficient $C$ of the QMS. In principle, this is measured by applying a constant pressure of gas at room temperature to the QMS vacuum chamber, and measuring the QMS count rate while monitoring the pressure with a Bayard-Alpert gauge. This procedure suffers from several problems. First, even taking into account gas correction factors, the specified accuracy of our Bayard-Alpert gauge is at best 20\,\%. Second, since a titanium sublimation pump attached to the system cannot easily be turned off, a constant input stream of gas is needed to maintain a constant pressure. It is thus not clear if the pressure at the QMS is equal to the pressure at the Bayard-Alpert gauge. Finally, depending on the molecule, fragmentation in the QMS and in the Bayard-Alpert gauge can lead to a substantial partial pressure of fragments in the vacuum chamber which distorts the result. To reduce the third effect, we measure the sensitivity coefficient of the QMS for oxygen and nitrogen and extrapolate the value for formaldehyde based on ionization cross sections and fragmentation probabilities for electron impact ionization from the literature~\cite{Nistelectron,Nistchem}. The sensitivity coefficient for formaldehyde obtained from oxygen differs from the one obtained from nitrogen by about 15\,\%, which we attribute to the sources of error identified above.

Obtaining the overall error for the number of molecules exiting the trap is difficult since the main source of error, from the sensitivity coefficient of the QMS, is hard to quantify, as discussed above. As a conservative value, we estimate that the number of molecules exiting the trap is determined to within at least a factor of two. We note that when comparing different ensembles in the trap, the effect of the sensitivity coefficient cancels, so that relative populations can be determined much more accurately.

\end{document}